\pdfoutput=1
\documentclass[a4paper,11pt, oneside]{article}
\usepackage[dvipsnames]{xcolor}
\usepackage{amsmath}
\usepackage[a4paper, margin=0.7in]{geometry}
\usepackage{accents}
\usepackage{multicol}
\newlength{\dhatheight}

\newcommand{\realnumbers}{\rm I\!R}
\newcommand{\argmax}{\mathop{\mathrm{argmax}}}
\usepackage[english]{babel}
\usepackage[utf8]{inputenc}
\usepackage{url}
\usepackage[pdftex]{graphicx}
\usepackage{abstract}
\usepackage{mathtools}
\usepackage{float}
\usepackage{footnote}
\usepackage{etoolbox}
\newcommand{\authorblock}[1]{\begin{tabular}{@{}c@{}}#1\end{tabular}}
\renewcommand{\vec}{\textbf}
\usepackage[switch,columnwise]{lineno}
\usepackage{hyperref}
\usepackage{tikz}
\usetikzlibrary{shapes.geometric, arrows}
\usepackage{varwidth}
\usetikzlibrary{decorations.pathreplacing}
\tikzstyle{arrow}=[draw, -latex]
\tikzstyle{current} = [rectangle, minimum width=1cm, minimum height=1cm,text centered, draw=black, execute at begin node={\begin{varwidth}{1cm}}, execute at end node={\end{varwidth}}]
\tikzstyle{normal} = [rectangle, minimum width=1cm, minimum height=1cm,text centered, draw=black, execute at begin node={\begin{varwidth}{1cm}}, execute at end node={\end{varwidth}}]
\tikzstyle{c1} = [rectangle, fill=gray, fill opacity=0.3, text opacity=1, minimum width=1cm, minimum height=1cm,text centered, draw=black, execute at begin node={\begin{varwidth}{1cm}}, execute at end node={\end{varwidth}}]
\tikzstyle{c2} = [rectangle, minimum width=1cm, minimum height=1cm,text centered, draw=black, execute at begin node={\begin{varwidth}{1cm}}, execute at end node={\end{varwidth}}]
\tikzstyle{c3} = [rectangle, fill=red, fill opacity=0.3, text opacity=1, minimum width=1cm, minimum height=1cm,text centered, draw=black, execute at begin node={\begin{varwidth}{1cm}}, execute at end node={\end{varwidth}}]
\tikzstyle{c4} = [rectangle, fill=red, minimum width=1cm, minimum height=1cm,text centered, draw=black, execute at begin node={\begin{varwidth}{1cm}}, execute at end node={\end{varwidth}}]
\tikzstyle{c5} = [rectangle, fill=green, minimum width=1cm, minimum height=1cm,text centered, draw=black, execute at begin node={\begin{varwidth}{1cm}}, execute at end node={\end{varwidth}}]
\AtEndDocument{%
    \bibliography{evolutionary}
}
\title{Evolutionary algorithms for hyperparameter optimization in machine learning for application in high energy physics}
\author{\begin{tabular}{c@{\qquad}c}
  \authorblock{Tani, Laurits$^1$\\ \texttt{laurits.tani@cern.ch} } &
  \authorblock{Rand, Diana$^1$ \\ \texttt{diana.rand@cern.ch}} \\[\bigskipamount]
  \authorblock{Veelken, Christian$^1$ \\ \texttt{Christian.Veelken@cern.ch}} &
  \authorblock{Kadastik, Mario$^1$ \\ \texttt{Mario.Kadastik@cern.ch}}
\end{tabular}}
\date{%
    {\footnotesize $^1$\textit{National Institute Of Chemical Physics And Biophysics (NICPB), Akadeemia tee 23, 12618 Tallinn, Estonia}\\}
    \vspace{0.5cm}
    \today
}

\begin{document}

\maketitle
    The analysis of vast amounts of data constitutes a major challenge in modern high energy physics experiments. Machine learning (ML) methods, typically trained on simulated data, are often employed to facilitate this task. 
    Several choices need to be made by the user when training the ML algorithm. In addition to deciding which ML algorithm to use and choosing suitable observables as inputs, users typically need to choose among a plethora of algorithm-specific parameters. We refer to parameters that need to be chosen by the user as hyperparameters. These are to be distinguished from parameters that the ML algorithm learns autonomously during the training, without intervention by the user. The choice of hyperparameters is conventionally done manually by the user and often has a significant impact on the performance of the ML algorithm.
    In this paper, we explore two evolutionary algorithms: particle swarm optimization (PSO) and genetic algorithm (GA), for the purposes of performing the choice of optimal hyperparameter values in an autonomous manner. Both of these algorithms will be tested on different datasets and compared to alternative methods.

\begin{multicols}{2}
\section{Introduction}
\label{intro}
  Owing to the large amount of data recorded by contemporary high energy physics (HEP) experiments, the analysis of data relies on powerful computing facilities. Machine learning (ML) methods are used extensively to aid the data analysis~\cite{albertsson2018machine,whiteson2009machine}.
  Boosted decision trees (BDTs)~\cite{roe2005boosted} and artificial neural networks (ANNs)~\cite{jain1996artificial} are commonly used in HEP experiments.
  Even though these methods may aid the data analysis task significantly, their usage in practical HEP applications is not trivial. This is because, in order to achieve optimal results, a set of parameters, referred to as hyperparameters in the literature~\cite{claesen2015hyperparameter}, need to be chosen by the user, depending on the given task and data.

  The subject of this paper is to describe two evolutionary algorithms~\cite{zitzler2000comparison}, which allow to find a set of optimal hyperparameters in an autonomous manner.
  The evolutionary algorithms studied in this paper are particle swarm optimization (PSO)~\cite{kennedy1995particle} and genetic algorithm (GA)~\cite{holland1992genetic}.

  The task of finding optimal hyperparameter values can be recast as function maximization.
  One considers a mapping from a point $h$ in hyperparameter space $\mathcal{H}$ to a "score" value $s(h)$, which quantifies the performance of the ML algorithm for a given task. Using a suitable encoding for hyperparameters of non-floating-point type, the hyperparameter space $\mathcal{H}$ can be taken to be the Euclidean space $\realnumbers^{N}$, with $N$ denoting the number of hyperparameters.
  Formally, the optimal hyperparameters, denoted by the symbol $\hat{h}$, are those that satisfy the condition:
  \begin{equation}
  \hat{h} = \argmax_{h \in \mathcal{H}} \, s(h) \, ,    
  \end{equation}
  where $s: \, \mathcal{H} \mapsto \realnumbers$ refers to the objective function that maps a point $h$ in $\mathcal{H}$ to a score $s(h)$.
  Recasting the hyperparameter optimization task as a function maximization problem allows to evaluate the performance of the PSO and GA on reference problems on function maximization from literature, as well as to compare their performance with alternative methods.

  The paper is organized as follows: 
  In Sections \ref{sec:PSO} and \ref{sec:GA}, we describe the PSO and GA, respectively.
  In Section \ref{sec:data}, we apply both evolutionary algorithms to a well-known function minimization problem from the literature, based on the Rosenbrock function~\cite{shang2006note}, as well as to a typical data analysis task from the domain of HEP, the "ATLAS Higgs boson machine learning challenge"~\cite{adam2015higgs}.
  We conclude the paper with a summary in Section~\ref{sec:summary}.

\section{Particle Swarm Optimization}\label{sec:PSO}

  Particle swarm optimization (PSO) \cite{kennedy1995particle} represents a computational method for optimizing continuous nonlinear functions. The method is effective for optimizing a wide range of functions.
  In common with other evolutionary algorithms, such as the GA, the PSO method is inspired by nature.

  %%%%%%%%%%%%%%%%%%%%%%%%%%%%%%%%%%%%%%%%%%%%%%%%%%%%%%%%%%%%%%%%%%%%%%%%%%%%%%%

  As the name of the method implies, the maximization of the objective function by the PSO is performed by a \emph{swarm of particles}. The particles traverse the hyperparameter space $\mathcal{H}$, with the position of each particle representing one set of hyperparameters $h$.
  Having a swarm of particles allows the exploration of multiple points in the space $\mathcal{H}$ in parallel, thereby allowing for a highly parallel implementation of the PSO algorithm on a computer.
  The evolution of the particle swarm proceeds in iterations denoted by the letter $k$.
  In each iteration a new position $\textbf{x}^{k+1}_{i}$ is computed for each particle $i$ according to the relation:

  \begin{equation}\label{eq:newLoc}
      \vec{x}_{i}^{k+1} = \vec{x}^k_{i} + w\cdot \vec{p}^{k}_{i} + \vec{F}^{k}_{i}
  \end{equation}

  where $\vec x^{k}_{i}$ denotes the current position of the particle and $\vec p_{i}^{k}$ its \emph{momentum}.
  The momentum term $w\cdot\vec p_{i}^{k}$ represents the inertia for particles to change their direction when traversing the space $\mathcal{H}$.
  The symbol $\vec{F}_{i}^{k}$ represents an attractive \emph{force}, which has the effect for particles to move towards previously discovered extrema of the objective function.
  The momentum term causes a tendency for the particles to continue moving in their current direction, past the previously found extrema.
  This behaviour increases the exploration of the hyperparameter space $\mathcal{H}$  and is found to improve performance \cite{kennedy1995particle}.
  The coefficient $w$ is refered to as \emph{inertial weight} in the literature \cite{shi1998parameter}, though the term \emph{damping weight} might be actually more descriptive as suggested in reference \cite{eberhart1998comparison}.

  Our implementation of the PSO algorithm distinguishes between the personal best location $\hat{x}_{i}^{k} = \{x \in \mathcal{H} \wedge \hat{x}_{i}^{k}=x_{i}^{k'}$ for $k'\leq k \wedge s(x_{i}^{k}) \leq s(\hat{x}_{i}^{k}) \forall k' \leq k\}$ and the best known global extremum $\hat{\hat{x}}^{k} = \argmax \{\hat{x}_{i}^{k}\}$:

  \begin{equation}
      \vec F_{i}^{k} = c_{1} \cdot r_{1} \cdot (\hat{\vec{x}}_{i}^{k} - \vec{x}_{i}^{k}) + c_{2} \cdot r_{2} \cdot (\hat{\hat{\vec{x}}}^{k} -\vec{x}_{i}^{k}).
  \end{equation}

  The coefficients $c_{1}$ and $c_{2}$ are referred to as the \emph{cognitive} and the \emph{social} weights in the literature \cite{van2006study}, and the symbols $r_{1}$ and $r_{2}$ represent random numbers, which are drawn from an uniform distribution in the interval [0,1]. The known global extremum $\hat{\hat{x}}^{k}$ for each particle is updated at each iteration by propagating the personal best location of a subset of the population, referred to as \emph{info}. The number of particles in this subset is denoted by $N_{info}$. Restricting the computation of $\hat{\hat{x}}^{k}$ to a subset of particles helps to avoid premature convergence of the swarm to a local minimum.

  We choose the coefficients $c_{1}$ and $c_{2}$ to be equal to 2, such that the particles move past their previously found target about half of the time if the inertial weight $w$ would be negligible \cite{kennedy1995particle}.

  After each iteration the momenta are updated according to the rule:

  \begin{equation}
      \vec{p}^{k+1}_{i} = \vec{x}_{i}^{k+1} - \vec{x}^{k}_{i}
  \end{equation}

  The positions $\vec{x}^{0}_{i}$ of all particles $i$ are initialized randomly within the hyperparameter space $\mathcal{H}$, while all momenta $\vec{p}_{i}^{0}$ are randomly initialized whithin one quarter of the range of each hyperparameter.

  The relation between the inertial weight \textit{w} and the size of the coefficients $c_{1}$ and $c_{2}$ controls the influence of global (wide-ranging) versus local (nearby) exploration abilities of the particles.
  A larger inertial weight \textit{w} allows the particles to move into unexplored regions of the hyperparameter space $\mathcal{H}$, whereas a small value of \textit{w} causes the particle to hone in on local and global extrema found previously \cite{shi1998parameter}.

   A suitable selection of \textit{w} can provide a balance between global and local exploration abilities and thus require fewer iterations on average to find the optimum \cite{shi1998parameter}.
  As discussed in Ref. \cite{eberhart1998comparison}, one may expect the performance of the PSO algorithm to be improved if one sets the inertial weight \textit{w} to a large value for the first iterations of the PSO algorithm and gradually reduces $w$ as the swarm evolves.
  Doing this allows the particles to explore the hyperparameter space $\mathcal{H}$ as fast as possible.
  By gradually reducing the value of \textit{w} during subsequent iterations, when the approximate location of the extremum has been established, one switches smoothly from the global exploration to a local exploration, thus improving on the accuracy of the found extrema.
  The idea is analogous to the gradual reduction of the temperature parameter in simulated annealing \cite{eberhart1998comparison}.

  Each time the position of a particle would move outside the bounds of the hyperparameter space $\mathcal{H}$, the position of the particle is set to the boundary value and its momentum is set to zero, thereby reducing the probability that the same particle moves against the boundary again in the next iteration.

\section{Genetic Algorithm}\label{sec:GA}
    The second evolutionary algorithm considered in this paper, the genetic algorithm (GA),  is motivated by the concept of natural selection \cite{holland1992genetic}.
    The GA maintains a population of possible solutions to the optimization problem, which evolve through multiple generations in order to produce the best solution.

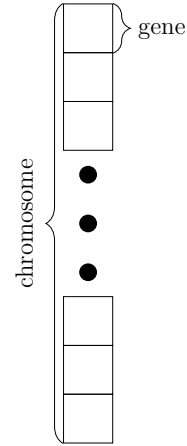
\begin{figure}[H]
    \centering
    \resizebox{0.4\columnwidth}{!}{
        \begin{tikzpicture}
            \node (i0) [current] {};
            \node (i) [current, below of=i0] {};
            \node (ii) [current, below of=i] {};
            \node (start) [right of=i] {};
            \node (p1) [fill, circle, minimum size=0.01cm, below of=ii] {};
            \node (p2) [fill, circle, minimum size=0.01cm, below of=p1] {};
            \node (p3) [fill, circle, minimum size=0.01cm, below of=p2] {};
            \node (iv) [current, below of=p3] {};
            \node (v) [current, below of=iv] {};
            \node (vi) [current, below of=v] {};
            \draw [decorate,decoration={brace, mirror, amplitude=10pt, raise=-2cm},rotate=270] ([yshift=-2cm, xshift= 0cm]i0.north west) -- ([yshift=-2cm, xshift=0cm]vi.south west) ;

            \draw [decorate,decoration={brace, amplitude=10pt, raise=-2cm},rotate=270] ([yshift=2cm, xshift= 0cm]i0.north east) -- ([yshift=2cm, xshift=0cm]i0.south east) ;
%%%%%%%%%%%%%%%%%%%%%%%%%%%%%%%%%%%%%%%%%%%%%%%%%%

            \node[label={[label distance=cm,text depth=-1ex,rotate=90]center:\Large chromosome}, left of=p2, xshift=-0.3cm] {};
            \node[label={[label distance=cm,text depth=-1ex,rotate=0]center:\Large gene}, right of=i0, xshift=.5cm, yshift=0.cm] {};
        \end{tikzpicture}
        }
        \caption{A chromosome consisting of genes.}
        \label{fig:ga-genes}
    \end{figure}

    Each possible solution is referred to as a \textit{chromosome}.
    Each chromosome (see Fig. \ref{fig:ga-genes}) represents one point in the hyperparameter space $\mathcal{H}$.
    Having multiple chromosomes allows the GA to explore multiple solutions in parallel.

    The number of genes in a chromosome matches the dimension of the hyperparameter space $\mathcal{H}$.

    The evolution towards the best solution is iterative.
    Each iteration corresponds to one generation in the evolution of all chromosomes and consists of 3 distinct stages: the selection of parents, the crossover of the genes, and the mutation.
    
    The selection of parents is performed using the tournament method \cite{yang1997structural,razali2011genetic}.
    In each tournament a certain number of chromosomes compete to be selected as a parent for the next generation.
    The number of chromosomes participating in each tournament is denoted by the symbol $N_{tour}$.
    The participants are drawn from the population of chromosomes at random and are ranked in order of decreasing score $s(h)$.
    The participant with the highest score is selected as a parent with the probability $P_{tour}$.
    In case the chromosome with the highest score is not selected, the chromosome with the second highest score gets selected, again with the probability $P_{tour}$, and so on.
    The tournament ends when two chromosomes are selected in this way to be the parents.
    
    A larger value of $N_{tour}$ has the effect that the chromosome with a low score $s(h)$ has a smaller chance to be selected as the parent for the next generation, because there is a high probability that a chromosome with a better score participates in the same tournament.
    A smaller value of $N_{tour}$ has the opposite effect.
    New tournaments are started until a sufficient number of pairs of parents are selected to produce the chromosomes for the next generation.
    
\begin{figure}[H]
            \centering
            \resizebox{0.8\columnwidth}{!}{
            \begin{tikzpicture}
                \node (i) [c5] {$h_{11}$};
                \node (ii) [c5, below of=i] {$h_{12}$};
                \node (iii) [c2, below of=ii] {$h_{13}$};
                \node (iv) [c2, below of=iii] {$h_{14}$};
                \node (v) [c5, below of=iv] {$h_{15}$};
                \node (vi)[c5, below of=v] {$h_{16}$};

                \node (ci) [c5, xshift=3cm, right of=i] {$h_{11}$};
                \node (cii) [c5, xshift=3cm, right of=ii] {$h_{12}$};
                \node (ciii) [c4, xshift=3cm, right of=iii] {$h_{23}$};
                \node (civ) [c4, xshift=3cm, right of=iv] {$h_{24}$};
                \node (cv) [c5, xshift=3cm, right of=v] {$h_{15}$};
                \node (cvi)[c5, xshift=3cm, right of=vi] {$h_{16}$};

                \node (ri) [c2, xshift=7cm, right of=i] {$h_{21}$};
                \node (rii) [c2, xshift=7cm, right of=ii] {$h_{22}$};
                \node (riii) [c4, xshift=7cm, right of=iii] {$h_{23}$};
                \node (riv) [c4, xshift=7cm, right of=iv] {$h_{24}$};
                \node (rv) [c2, xshift=7cm, right of=v] {$h_{25}$};
                \node (rvi)[c2, xshift=7cm, right of=vi] {$h_{26}$};

                \node (c11)[right of=ii, xshift=-0.5cm, yshift=0.5cm]{};
                \node (c21)[left of=ci, xshift=0.5cm, yshift=-0.5cm]{};

                \node (c32)[left of=riii, xshift=0.5cm, yshift=-0.5cm]{};
                \node (c22)[right of=ciii, xshift=-0.5cm, yshift=-0.5cm]{};

                \node (c13)[right of=vi, xshift=-0.5cm, yshift=0.5cm]{};
                \node (c23)[left of=cv, xshift=0.5cm, yshift=-0.5cm]{};

                \draw [arrow, yshift=2cm] (c11) -- (c21);
                \draw [arrow, yshift=5cm] (c32) -- (c22);
                \draw [arrow, yshift=5cm] (c13) -- (c23);

                \node[label={[label distance=cm,text depth=-1ex,rotate=0]center:\Large parent1}, below of=vi, xshift=-0.3cm] {};
                \node[label={[label distance=cm,text depth=-1ex,rotate=0]center:\Large parent2}, below of=rvi, xshift=-0.3cm] {};
                \node[label={[label distance=cm,text depth=-1ex,rotate=0]center:\Large offspring}, below of=cvi, xshift=-0.3cm] {};
            \end{tikzpicture}
            }
            \caption{Possible outcome of a 2-point crossover of two parents in case of 6-dimensional hyperparameter space $\mathcal{H}$, where $h_{11}$ denotes the first hyperparameter of the first parent, $h_{12}$ the second hyperparameter of the first parent, etc.}
            \label{fig:ga-crossover}
            \end{figure}
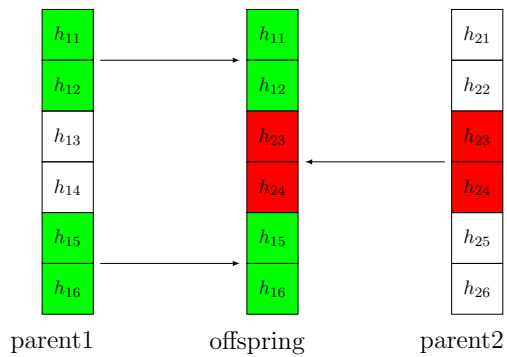

    The chromosomes of two parents produce one new chromosome for the next generation by means of crossover \cite{spears1991analysis,de1992formal}.
    We use \emph{k-point crossover} in which the chromosomes of both parents are cut at $k$ points ($N_{cross}$ refers to the number of points, to avoid using the same symbol as for the number of iterations) and the chromosomes of the offspring are produced by randomly choosing chromosome segments from either parent (see Fig. \ref{fig:ga-crossover}).

    The chromosomes of the offspring that are obtained by the crossover operation are subject to \emph{mutation} \cite{holland1992genetic}, which aims to increase the diversity of the population, thereby allowing to explore domains in the hyperparameter space $\mathcal{H}$ not populated by chromosomes from the parent generation.
    Mutation also helps to avoid the population to get stuck in local minima.
    
    In our implementation of the GA, the mutation of chromosomes is performed by adding a random number, drawn from a normal distribution with a mean of zero and a given width, to each gene.
    A high mutation rate has the effect of turning the GA into a random search.
    We avoid this effect by linearly decreasing the width of the normal distribution each iteration, with the initial width corresponding to a quarter of the maximum range of a given hyperparameter and the final width corresponding to zero.

    Our implementation of the GA uses the concept of \emph{elitism} \cite{bhandari1996genetic}.
    Elitism means that the algorithm preserves a certain number of the best performing chromosomes within the population and passing the parent chromosomes on to the next generation together with their offspring.
    Elitism is found to improve the convergence toward an optimal solution.
    The number of parent chromosomes preserved in this manner is denoted by the symbol $N_{elite}$.
    
    The convergence is further enhanced by \emph{culling} \cite{baum2001genetic}, which means that we discard a certain number of chromosomes with the lowest score among the population before selecting the parents for the next generation.
    The number of parent chromosomes discarded in this way is referred to using the symbol $N_{cull}$.
    For each chromosome discarded by culling, we create a new chromosome with randomly initialized hyperparameter values to replace the one discarded.

    Our implementation of the GA further allows to evolve groups of chromosomes in \emph{subpopulations} \cite{tanese1989distributed}.
    The number of subpopulations is denoted by the symbol $N_{subpop}$.
    The selection of parents is restricted to the chromosomes from the same subpopulation for the first $N^{generations}_{subpop}$ iterations of the algorithm.
    For the remaining iterations, the chromosomes from different subpopulations are allowed to mix freely.

\section{Performance}\label{sec:data}
    The performance of both evolutionary algorithms, PSO and GA, is evaluated on two tasks: on the Rosenbrock function, which provides an example for a difficult function minimization problem, and on the ATLAS Higgs boson machine learning (ML) challenge, as a typical application of ML methods in HEP.

    \subsection{Rosenbrock function}\label{sec:rosenbrock}
        The Rosenbrock function \cite{shang2006note,kok2009locating}  represents a well-known trial function for evaluating the performance of function minimization algorithms.
        The function is defined as:

        \begin{equation}\label{eq:rosenbrock}
            R(x, y) = (a - x)^2 + b(y - x^2)^2,
        \end{equation}

        \noindent where the $a$ and $b$ are constants.
        
        The Rosenbrock function has a global minimum at $(x, y) = (a, a^2)$. We chose to study the Rosenbrock function for the case $a = 1$ and $b = 100$. For the chosen values of $a$ and $b$, the global minimum is located at the position $(x, y) = (1, 1)$, and the function value at the minimum is $R(1, 1) = 0$.

        \begin{figure}[H]
            \includegraphics[width=\linewidth]{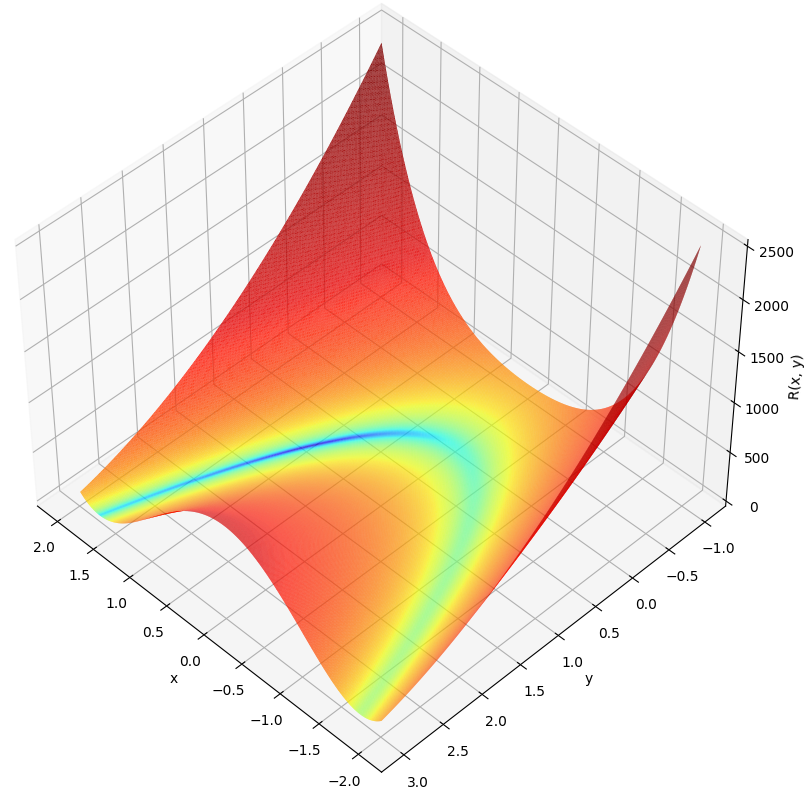}
            \caption{The Rosenbrock function in a region around its global minimum, located at the position $(x, y) = (1, 1)$.}
            \label{fig:rosenbrock}
        \end{figure}

        The challenge in finding the global minimum of the Rosenbrock function is that the function varies slowly along a curved valley, while rising steeply in direction orthogonal to the valley.
        Function minimization algorithms hence need to closely track the location of the valley. Fig. \ref{fig:rosenbrock} illustrates the Rosenbrock function in the region around the global minimum.
        
        For the purpose of evaluating the performance of the PSO and GA, we treat the minimization of $R(x,y)$ as function of $x$ and $y$ as a two dimensional hyperparameter optimization problem, identifying x and y with the first and second hyperparameter respectively.
        The position of particles in case of the PSO algorithm and the value of the chromosomes in the case of GA are initialized within the range $[-500, +500] \times [-500, +500]$ and are enforced to stay within this range during the evolution of both algorithms.

        \subsubsection{Stopping Criteria}\label{sec:stop-crit}
            In order to limit the computing time, we define a criterion when to stop the training of the PSO and GA.
            We use two criteria for this purpose and terminate the evolution when either criterion is fulfilled.
            The first criterion is an upper limit on the number of iterations, denoted by the symbol $N_{iter}^{max}$.
            Additionally, we terminate the evolution once the algorithm has found a point (x, y) for which $R(x, y) < 10^{-3}$.

        \subsubsection{Optimization methods}

            We compare the performance of the PSO and GA for finding the minimum of the Rosenbrock function with three alternative methods, the gradient descent algorithm \cite{ruder2016overview}, and two naive methods for choosing the hyperparameters, to which we refer to as "grid search" and "random guessing". The latter two serve as a cross-check.
            One would expect of course that evolutionary algorithms such as the PSO and GA outperform the naive methods.

        \paragraph{(Modified) gradient descent}
        
            We have modified the gradient descent (GD) algorithm in order to improve its performance on the Rosenbrock function.
            The issue is that the unmodified GD algorithm often 'zig-zags' from one side of the valley to the other, causing the algorithm to progress very slowly in the direction along the valley, towards the global minimum \cite{andrychowicz2016learning}.
            To prevent this 'zig-zag' behaviour and improve the convergence of the algorithm, we have modified the GD algorithm in the following way: at each iteration, the algorithm determines the direction of the steepest descent by numerical evaluation of the gradient at a given point $h^{k}$.
            The new position $h^{k+1}$ is computed according to:
            
            \begin{equation}
                h^{k+1} = h^k  + \delta \cdot \frac{\nabla h^{k}}{|\nabla h^{k}|},
            \end{equation}
            
            where the term $\frac{\nabla h^{k}}{|\nabla h^{k}|}$ represents the direction of the steepest descent and the step size $\delta$ represents the parameter of the algorithm.
            
            Rather than moving immediately to the new position $h^{k+1}$, the modified GD algorithm computes the value of the objective function  $s$ at the new position $s(h^{k+1})$.
            
            It then compares the actual decrease of the objective function, $s(h^{k+1}) - s(h^{k})$, with the expected decrease, given the expression $\delta \cdot \frac{\nabla h^{k}}{|\nabla h^{k}|} \cdot \nabla h^{k}$. In case $s(h^{k+1}) - s(h^k) > 2 \cdot \delta \cdot \frac{\nabla h^{k}}{|\nabla h^{k}|} \cdot \nabla h^{k}$, we conclude that the step size $\delta$ is too large and needs to be reduced in order to avoid this 'zig-zag' behaviour.

            In our implementation, we successively reduce the step size by a factor of two until the condition is satisfied.
            The algorithm then moves to the new position, the initial step size is restored, and the algorithm recomputes the gradient at the new position for the next iteration.
            
            We choose the number of iterations for the GD algorithm to be $10^{6}$ and the initial step size $\delta$ to be $10^{-2}$.

        \paragraph{Grid search}
            
            This is a widely used hyperparameter optimization method available for example in the package scikit-learn \cite{scikit-learn}. This method is based on choosing $N^d$ grid points in each dimension $d$ of the hyperparameter space $\mathcal{H}$, evaluating the objective function $s$ for all $\prod_{d=1}^{N} N^{d}_{grid}$ combinations of grid points, and selecting the best performing combination. The same number of evaluations of the objective function, $\prod_{d=1}^{N} N^{d}_{grid}$, is chosen to be the same as for the other algorithms, in order to compare all algorithms for the same time usage of computing time.
            Here we assume that the evaluation of the objective function consumes the majority of the computing time and the computations internal to the PSO and GA are negligible in comparison.
            We believe this assumption represents a very good approximation for practical approximations of these methods in HEP, discussed in the introduction, where one evaluation of $s$ corresponds to one training of a ML algorithm.
            For the Rosenbrock function minimization task, we choose $N_{grid}^1 = N_{grid}^2 = 10^3$ grid points for each of the two dimensions, equidistantly within the interval [-500, +500] in each dimension.

        \paragraph{Random Guessing}
            In the random guessing (RNG) method, we draw a total of $N_{p} = 10^{6}$ points in the hyperparameter space $\mathcal{H}$ at random, sampling from a uniform distribution within the range $[-500, +500] \times [-500, +500]$. The point corresponding to the minimum of the objective function $s$ over the set of these points is selected as the best-performing point of the RNG method.
            The number of points $N_{p}$ is chosen such that the function is evaluated the same number of times for the RNG method as for the PSO, GA, GD and GS methods.

        \paragraph{Particle swarm optimization}
            The same maximal number of $10^6$ evaluations of the objective function $s$ were used for the PSO, by setting the number of particles in the swarm to 100 and the maximum number of iterations to $10^4$.
            The evaluation of the PSO was terminated before reaching the maximum number of iterations in case the global minimum $s(\hat{\hat{x}}^k)$ found by the PSO differed from the global minimum of the Rosenbrock function by less than $10^{-3}$.
            The coefficients $c_1$ and $c_2$ were chosen to be 2 and the inertial weight $w$ was chosen to linearly decrease from 0.8 to 0.4 as a function of iteration $k$. The number of informants $N_{info}$ was set to 7.

        \paragraph{Genetic Algorithm}
            The same maximum number of $10^6$ evaluations of the objective function were used for the GA, for which the number of chromosomes was chosen to be $10^4$ and the maximum number of iterations to be 100.
            The same threshold for early termination of $10^{-3}$ was chosen for the GA, as for the PSO.
            The early termination triggers once $s(h) < 10^{-3}$ for the hyperparameter values $h$ represented by any chromosome .
            The values of other parameters of the GA, used for the Rosenbrock function  minimization task, are given in Table \ref{tab:ga-settings-rosenbrock}.
            During the first iterations of the algorithm, when subpopulations are used, the parameters $N_{cull}$ and $N_{elites}$ amount to 10 and 5 respectively.

            \begin{table}[H]
            \begin{center}
                \caption[]{Parameters of the GA used for the Rosenbrock function minimization task.}
                \begin{tabular}{l | c}
                    Parameter & Value \\
                    \hline
                    $N_{tour}$ & 5\\
                    $P_{tour}$ & 0.4\\
                    $N_{cross}$ & 1\\
                    $P_{mutate}$ & 0.2 \\
                    $N_{subpop}^{generations}$ & 90 \\
                    $N_{subpop}$ & 5\\
                    $N_{cull}$ & 50\\
                    $N_{elite}$ & 25\\
                    \hline
                \end{tabular}
                \label{tab:ga-settings-rosenbrock}
            \end{center}
            \end{table}

        \subsubsection{Procedure for comparing different methods}
            Owing to the fact that the minima found by the GD, GS and RNG, PSO, GA methods depends on the values of random numbers that are used to initialize and/or evolve each algorithm, the performance of each method needs to be evaluated for a set of different 'trials', each trial using a different seed to produce a different sequence of random numbers.
        
        \subsubsection{Results}
        
\begin{figure}[H]
                \centering
                \includegraphics[width=\columnwidth]{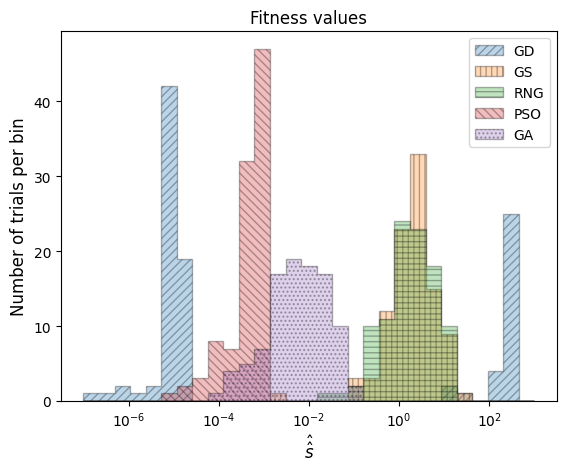}
                \caption{Distribution in $\hat{\hat{R}} = R(\hat{\hat{h}})$ of the Rosenbrock function at the minimum $\hat{\hat{h}}$ found in 100 different trials for the GD, GS, RNG, PSO and GA.}
                \label{fig:fitnesses}
            \end{figure}

            The distribution in $\hat{\hat{R}} = R(\hat{\hat{h}})$ at the minima $\hat{\hat{h}}$ found in 100 different trials is shown in Fig. \ref{fig:fitnesses}.
            Numerical values of the average $\bar{R}$ and of the width of the distribution, quantified by the standard deviation $\sigma_{R} = \sqrt{\frac{1}{99}\sum s}$, are shown in Table \ref{tab:stability-rosenbrock}.

            \begin{table}[H]
            \begin{center}
                \caption{Average value $\bar{R}$ and standard deviation $\sigma_{R}$ achieved by the GD, GS, RNG, PSO, and GA methods in the Rosenbrock function minimization task.}
                \begin{tabular}{c | c | c}
                    Method & $\bar{R}$ & $\sigma_{R}$ \\
                    \hline
                    GD & 85.85 & 143.7 \\
                    GS & 3.29 & 3.89 \\
                    RNG & 3.11 & 3.44 \\
                    PSO & 0.00057 & 0.00030 \\
                    GA & 0.0014 & 0.0021 \\
                    \hline
                \end{tabular}
                \label{tab:stability-rosenbrock}
            \end{center}
            \end{table}

            \paragraph{Discussion}
            One can see in Fig. \ref{fig:fitnesses} that the GD method performs extraordinarily well in about half of the trials, while in the other half it fails to get close to the minimum of the Rosenbrock function at all. The poor performance of the GD method in the latter trials is due to the cases where the particle moves so slowly along the valley of the Rosenbrock function that the maximum number of $10^6$ iterations is reached before the algorithm reaches the global minimum of the position $(x, y) = (1, 1)$.
            
            The PSO algorithm achieves the lowest value $\bar{R}$, outperforming all other methods on the Rosenbrock function minimization task, followed by the GA.
            The PSO and GA also exhibit the lowest standard deviation $\sigma_{R}$, which means that their performance is robust against variations in the random choice of starting positions across different trials.
            
            We remark that the early termination limited the average number of evaluations of the objective function to $\sim 7\cdot10^3$ for the PSO, while the early termination had little effect for the GA (as well as for the GD, GS, and RNG methods), which makes the performance of the PSO even more impressive. As expected, both evolutionary algorithms outperform all other methods.

    \subsection{The ATLAS Higgs boson machine learning challenge}
    
        The ATLAS Higgs boson machine learning challenge (HBC) \cite{adam2015higgs} represents a typical application of ML algorithms to the field of HEP.
        The task of the HBC is to obtain an optimal separation of the Standard Model (SM) Higgs boson $\rightarrow \tau\tau$ signal from the large SM background.
        The background consist of Drell-Yan production of Z bosons, the production of W bosons in association with jets, and top quark pair production.
        Samples of signal and background events are generated by Monte Carlo (MC) simulation.
        Events are selected in the $\tau\tau \rightarrow \mu \bar{\nu}_{\mu}\nu_{\tau} \tau_h \nu$ final state, where we use the symbol $\tau_{h}$ to denote the hadronic decay of a $\tau$ lepton.
        Background contributions arising from multijet production without associated production of bosons or top quark are neglected.
        
        In total 550 000 signal plus background events are provided by the organizers of the HBC, of which we use 80\% for training the ML algorithm and 20\% for testing the performance of the trained ML algorithm. We refer to the former as the train samples and to the latter as the test sample.
        
        We utilize a BDT to perform the separation of the Higgs boson signal from backgrounds.
        For the BDT implementation, we chose the  XGBoost package \cite{chen2016xgboost}.
        
        The objective function $s$ for the hyperparameter optimization represents an approximation for the sensitivity to discover the Higgs boson signal in a physics analysis at the Large Hadron Collider (LHC).
        The function $s$ was given by the organizers of the HBC and is referred to as the 'approximate mean significance' (AMS), which is defined by:
        
        \begin{equation}\label{eq:ams}
            AMS(\theta_{cut}) = \sqrt{2\cdot(s + b + b_r)\cdot ln [1 + \frac{s}{b + b_r}] - s} \, \, ,
        \end{equation}
        
        \noindent where $b$ denotes the amount of background and $s$ the amount of signal that passes a cut on the BDT output. The term $b_r$ is introduced as a regularization in order to reduce the effect of statistical fluctuations of $b$ and $s$, resulting from limited MC statistics (as discussed in Ref. \cite{adam2015higgs}).
        The value of $b_r$ was given by the organizers of the HBC and amounts to $b_r = 10$. The function $AMS(\theta_{cut})$, for $\theta_{cut} = 0.15$, is used as objective function for the BDT training.
        
        Even with the addition of the $b_r$ term, statistical fluctuations of the number of signal and background events passing the cut on the BDT output still causes a sizable difference between the AMS scores computed on the test and on the training sample.
        We find that the difference between test and training performance can be reduced and a higher AMS score on the test sample can be achieved if we use a modified version of Eq. (\ref{eq:ams}) as the objective function for the BDT training.
        We refer to the modified version of Eq. (\ref{eq:ams}) as d-AMS.
        The idea is to add a penalty term for the difference between the AMS scores on the test compared to the training sample, so that the BDT training (and the hyperparameter optimization) reduces this difference:
        
            \begin{equation}\label{eq:d-score}
                \begin{split}
                    d{\text -}AMS &= AMS_{test}\\
                    &- \kappa \cdot \mathrm{max}(0, [AMS_{test} - AMS_{train}])
                \end{split}
            \end{equation}
        
        \noindent where the coefficient $\kappa$ controls the strength of the penalty term.
        We find the choice $\kappa = 1.5$ to work well for a wide range of different ML applications that we tried.
        After the BDT training with a fixed $\theta_{cut} = 0.15$ has finished, the threshold $\theta_{cut}$ is optimized such that d-AMS attains its maximal value on the training sample.
        
        The PSO was evolved for a maximum of 7000 evaluations of the objective function, using a swarm of 70 particles and a maximum number of 100 iterations.
        The coefficients $c_1$ and $c_2$ were both chosen to be equal to 2 and the inertial weight $w$ was chosen to linearly decrease from 0.8 to 0.4 during the evolution of the PSO algorithm.
        
        For the GA, we used 70 chromosomes and a maximum number of 100 iterations.
        The values of the other parameters are given in Table \ref{tab:ga-parameters-higgs}.
        
        The XGBoost hyperparameters chosen and the default values for these parameters are given in Table \ref{tab:default-parameters}.
        The parameter \emph{num-boost-round} specifies the number of boosting iterations, corresponding to the number of trees in the BDT.
        The \emph{learning-rate} parameter controls the effect that trees added at a later stage of the boosted iterations have on the output of the BDT relative to the effect of trees added at an earlier stage.
        Small values of the \emph{learning-rate} parameter decrease the effect of trees added during the boosting iterations, thereby reducing the effect of boosting on the BDT output.
        The parameter \emph{max-depth} specifies the maximum depth of a tree.
        The parameter \emph{gamma} represents a regularization parameter, which aims to reduce overfitting. Large values of this parameter prevent the splitting of leaf nodes before the maximum depth of a tree is reached.
        The parameter \emph{min-child-weight} specifies the minimum number of events that is required in each leaf node.
        The parameter \emph{subsample} limits the number of training events that are used to grow each tree to a fraction of the full training sample. A value of this parameter smaller than one decreases overfitting.
        The parameter \emph{colsample-bytree} specifies the number of different features that are used in a tree.
        A value of one means that all features are considered for splitting leaf nodes, while a value smaller than one restricts the number of features that are used in a tree to a subset of all features. The purpose of this restriction is to reduce overfitting. The number of features considered for each tree are drawn at random, independently for each boosting iteration.
        
        The choice of all of these parameters typically represents a trade-off.
        Large values of the parameters \emph{num-boost-round}, \emph{learning-rate}, \emph{max-depth}, \emph{subsample}, and \emph{colsample-bytree} increase the complexity of the BDT, while large values of the parameters \emph{gamma}, and \emph{min-child-weight} have a regularizing effect.
        BDTs with a higher complexity in general perform better in separating signal from background on the training sample, but typically are also more susceptible to overfitting.

        \begin{table}[H]
        \begin{center}
            \caption{Default values of hyperparameters in the XGBoost package \cite{default_xgb}.}
            \begin{tabular}{c | c}
                Parameter & Default value\\
                \hline
                \emph{num-boost-round} & 10\\
                \emph{learning-rate} & 0.3\\
                \emph{max-depth} & 6\\
                \emph{gamma} & 0\\
                \emph{min-child-weight} & 1.0\\
                \emph{subsample} & 1.0\\
                \emph{colsample-bytree} & 1.0\\
                \hline
            \end{tabular}
            \label{tab:default-parameters}
        \end{center}
        \end{table}

        The performance of the PSO and GA is assessed by comparing the AMS scores achieved on the test sample by a BDT trained with the default hyperparameters and with hyperparameters obtained by the RNG method compared to the AMS scores of BDTs trained with optimized hyperparameter values found by the PSO and GA.

            \begin{table}[H]
            \begin{center}
                \begin{tabular}{l | c}
                    Parameter & Value \\
                    \hline
                    
                    $N_{tour}$ & 5 \\
                    $P_{tour}$ & 0.4 \\
                    $N_{cross}$ & 1 \\
                    $P_{mutate}$ & 0.2 \\
                    $N_{subpop}^{generations}$ & 90 \\
                    $N_{subpop}$ & 5\\
                    $N_{cull}$ & 40\\
                    $N_{elite}$ & 7\\
                    \hline
                \end{tabular}
                \caption[]{Parameters of the GA used for the ATLAS Higgs boson machine learning challenge.}
                \label{tab:ga-parameters-higgs}
            \end{center}
            \end{table}

        Two criteria are used to stop the evolution of the PSO and GA. The first criterion is the number of iterations $N_{iter}$.
        Additionally, we terminate the evolution once the variance between the positions of the particles in the PSO or between chromosomes in the GA is below a certain threshold.
        The variance is quantified by the \emph{compactness} (also known as the mean coefficient of variance), which is defined as:

        \begin{equation}\label{eq:cov}
            \mbox{compactness} = \frac{1}{N} \sum_{j=1}^{N} \frac{\sigma^j}{\bar{x}^j}
        \end{equation}
        
        with
        
        \begin{equation*}
             \sigma^j = \sqrt{\frac{1}{n-1} \sum_{i=1}^{n}(x_{i}^{j} - \bar{x}^{j})^2},
        \end{equation*}

        where $N$ denotes the number of hyperparameters, $n$ the number of particles or chromosomes, and $\bar{x}^{j}$ the mean value of the j-th hyperparameter over the population of particles or genes, respectively.
        
        A low value of the compactness means that the hyperparameters of different particles or genes are very similar, indicating that the PSO or GA has converged to a single point in the hyperparameter space $\mathcal{H}$.

        \subsection{Results}

            The optimal values of the hyperparameters obtained with the RNG method, the PSO and the GA are given in Table \ref{tab:opt-higgs-parameters}.
            In Table \ref{tab:higgs-scores-comparison} we compare the AMS scores obtained for these hyperparameter values to the AMS scores obtained with the default values of hyperparameters defined in the XGBoost package.
            The performance is evaluated for two samples of events, referred to as the \emph{public} and \emph{private} leaderboard samples.
            Both samples are provided by the organizers of the HBC and contain signal and background events that overlap with neither the test nor the train sample.
            
            The performance achieved by the PSO and GA are very similar and about 12-13\% higher than the performance obtained using the default values of hyperparameters. The results of the BDT trained using the hyperparameters obtained by the RNG method are similar to those obtained by the PSO and GA.
            Comparing the PSO and GA optimized hyperparameters, we find that all except \emph{num-boost-round} and \emph{learning-rate} parameters have similar values.
            The value of the \emph{num-boost-round} parameter optimized by the GA is higher by about a factor of 2.9, while the value of the \emph{learning-rate} parameter is lower by a factor of 3.5. The fact that the \emph{learning-rate} parameter decreases by a factor that is similar to the increase of the \emph{num-boost-round} parameter is not surprising: using a large number of trees and a lower learning rate has about the same effect as using a lower number of trees and a higher learning rate.
            The product of the \emph{num-boost-round} and \emph{learning-rate} parameters is more similar, differing only by a factor of 1.2 between the PSO and GA.
            The situation is different for the \emph{colsample-bytree} parameter. It has a small effect on the d-AMS and AMS scores and is hence only loosely constrained during the hyperparameter optimization.
            
            The parameters obtained by the RNG method are more different - only the \emph{max-depth} parameter is very similar to those found by PSO and GA. Having \emph{min-child-weight} roughly two times smaller than it was found for PSO and GA means making the model more prone to overfitting. However, this effect is overcome by having the product of the anti-correlated pair, \emph{num-boost-round} and \emph{learning-rate}, two times smaller from the ones obtained by PSO and GA, thus the model less susceptible to overfitting. Furthermore having three to four times smaller \emph{gamma} helps the model generalize even further. Again the effect of \emph{colsample-bytree} had negligible effect on the d-AMS and AMS scores.

            \begin{table}[H]
            \begin{center}
                \caption{Hyperparameter values obtained by the RNG, PSO and the GA for the ATLAS Higgs boson machine learning challenge.}
                \begin{tabular}{c | c | c | c}
                    Parameter & RNG & PSO & GA\\
                    \hline
                    \emph{num-boost-round} & 295 & 153 & 451\\
                    \emph{learning-rate} & 0.062 & 0.300 & 0.085\\
                    \emph{max-depth} & 5 & 4 & 5\\
                    \emph{gamma} & 0.98 & 3.86 & 2.99\\
                    \emph{min-child-weight} & 173 & 323.6 & 442.2\\
                    \emph{subsample} & 0.83 & 0.830 & 0.907\\
                    \emph{colsample-bytree} & 0.7 & 1.0 & 0.3\\
                    \hline
                \end{tabular}
                \label{tab:opt-higgs-parameters}
            \end{center}
            \end{table}

                \begin{table}[H]
                \caption{Performance of BDTs trained using the optimal values of the hyperparameters obtained by the PSO and by the GA compared to a BDT trained using the default values of hyperparameters in the XGBoost package \cite{chen2016xgboost} and with hyperparameters obtained by RNG, for the ATLAS Higgs boson machine learning challenge.}
                \label{tab:higgs-scores-comparison}       % Give a unique label
                % For LaTeX tables use
                \begin{tabular}{cccc}
                \hline\noalign{\smallskip}
                Method & $\theta_{cut}$ & \begin{tabular}{@{}c@{}}AMS score \\ Public leaderboard\end{tabular}  &\begin{tabular}{@{}c@{}}AMS score \\ Private leaderboard\end{tabular} \\  \\
                \noalign{\smallskip}\hline\noalign{\smallskip}
                Default & 0.175 & 3.170 & 3.200\\
                RNG & 0.152 & 3.620 & 3.608 \\
                PSO & 0.134 & 3.628 & 3.655\\
                GA & 0.152 & 3.619 & 3.683\\
                \noalign{\smallskip}\hline
                \end{tabular}

                \end{table}

\section{Summary}\label{sec:summary}
  Two evolutionary algorithms, the particle swarm optimization (PSO) and the genetic algorithm (GA), for choosing an optimal set of hyperparameters in applications of machine learning (ML) methods to data analyses in high energy physics (HEP) have been presented.
  The performance of both methods have been studied for a difficult function minimization task (Rosenbrock function) and for a typical data analysis test in the field of HEP (Higgs boson machine learning challenge (HBC)).
  In the latter case, a boosted decision tree (BDT) has been used as ML algorithm. The PSO as well as the GA demonstrate their ability to find the optimal parameter value in the function minimization task.
  Compared to using the default values of hyperparameters, the optimization of the hyperparameter values improves the sensitivity of the data analysis, as quantified by the AMS score, by 12-13 \%.
  This improvement demonstrates that the optimization of hyperparameters is a worthwhile task for data analyses in the field of HEP.
  Randomly probing different hyperparameter sets and subsequently picking the best performing one showed similar performance to both PSO and GA. This can be attributed to the highly fluctuating hyperparameter space of this particluar example. For a highly structured hyperparameter space, the gain of using a more sophisicated method, like PSO or GA, will be much higher, as was shown by the Rosenbrock minimization problem. 
%   We expect both evolutionary algorithms to perform similarly well in case an artificial neural network (ANN) is used as ML algorithm.
  The optimization of the hyperparameters by the PSO and GA is fully automated and relieve the user from manual tuning of the hyperparameters.
\end{multicols}

\subsection*{Acknowledgements}
This work was supported by the Estonian Research Council grants PRG445 and PRG780. Also, we would like to thank Joosep Pata for constructive criticism of the manuscript.

\end{document}